\documentclass[aps,prb,floatfix,twocolumn,superscriptaddress]{revtex4-2}

\usepackage{mathrsfs}
\usepackage{graphicx}
\usepackage[english]{babel}
\usepackage{amsmath}
\usepackage{amssymb}
\usepackage{xcolor}
\usepackage{cancel}
\usepackage[normalem]{ulem}

\usepackage{relsize}
\usepackage{CJK}
\usepackage{dsfont}
\usepackage{bm}

\newcommand{\me}{\mathrm{e}}
\newcommand{\mi}{\mathrm{i}}

\newcommand{\dif}{\mathrm{d}}

\allowdisplaybreaks

\begin{document}
\title{Evolution of quantum geometric tensor of 1D periodic systems after a quench}

\author{Jia-Chen Tang}
\affiliation{School of Physics, Southeast University, Jiulonghu Campus, Nanjing 211189, China}

\author{Xu-Yang Hou}
\affiliation{School of Physics, Southeast University, Jiulonghu Campus, Nanjing 211189, China}

\author{Yu-Huan Huang}
\affiliation{School of Physics, Southeast University, Jiulonghu Campus, Nanjing 211189, China}

\author{Hao Guo}
\email{guohao.ph@seu.edu.cn}
\affiliation{School of Physics, Southeast University, Jiulonghu Campus, Nanjing 211189, China}
\affiliation{Hefei National Laboratory, University of Science and Technology of China, Hefei 230088, China}
\author{Chih-Chun Chien}
\email{cchien5@ucmerced.edu}
\affiliation{Department of physics, University of California, Merced, CA 95343, USA}

\begin{abstract}
We investigate the post-quench dynamics of the quantum geometric tensor (QGT) of 1D periodic systems with a suddenly changed Hamiltonian. The diagonal component with respect to the crystal momentum gives a metric corresponding to the variance of the time-evolved position, 
and its coefficient of the quadratic term in time is the group-velocity variance, signaling ballistic wavepacket dispersion.
The other diagonal QGT component with respect to time reveals the energy variance.
The off-diagonal QGT component features a real part as a covariance and an imaginary part representing a quench-induced curvature. Using the Su-Schrieffer-Heeger (SSH) model as an example, our numerical results of different quenches confirm that the post-quench QGT is governed by physical quantities and local geometric objects from the initial state and post-quench bands, such as the Berry connection, group velocities, and energy variance. Furthermore, the connections between the QGT and physical observables suggest the QGT as a comprehensive probe for nonequilibrium phenomena.
\end{abstract}

\maketitle

\section{Introduction}
The quantum geometric tensor (QGT) serves as a fundamental characterization of the local geometry of quantum states, measuring the distance between quantum states as well as the intrinsic curvature when the system varies with a set of parameters~\cite{QGTCMP80,BRODY200119,QGT10,KOLODRUBETZ20171}. For pure states, the QGT is a complex second-order tensor, with its real part corresponding to the Fubini-Study metric \cite{EGUCHI1980213} for measuring the distance between physical quantum states, and its imaginary part is proportional to the Berry curvature \cite{Simon83,Berry84} linking local geometry to global topological properties \cite{PhysRevLett.99.100603,PhysRevLett.117.045303,PhysRevB.90.165139,PhysRevA.92.063627,PhysRevB.105.045144,PhysRevLett.121.170401,PhysRevB.104.045103}. This structure makes the QGT essential in fields such as quantum information, condensed matter physics, and atomic, molecular, and optical (AMO) physics \cite{IG_Book,Bohm03,KOLODRUBETZ20171,QGTCMP80,cmp/1103904831,RevModPhys.82.1959,PhysRevResearch.3.L042018,PhysRevB.74.085308,PhysRevLett.72.3439,PhysRevB.103.014516, PhysRevB.103.205415, PhysRevB.102.155407,Park26,Resta26} (see Refs.~\cite{Yu25QGT,Gao26} for a review). Experimental platforms, including nitrogen-vacancy centers in diamonds~\cite{10.1093/nsr/nwz193}, superconducting qubits~\cite{PhysRevLett.122.210401}, photoemission spectroscopy~\cite{Kang2025}, and ultracold atoms~\cite{Yi23}, have utilized the QGT to probe physical observables through response functions and topological indicators \cite{PhysRevB.108.094508,PhysRevB.87.245103,PhysRevB.97.201117,PhysRevB.97.041108,PhysRevLett.121.020401,PhysRevB.97.195422,PhysRevLett.124.197002,PhysRevX.10.041041}. More applications of the QGT explore phenomena like quantum phase transitions \cite{PhysRevLett.99.100603}, orbital magnetic susceptibility \cite{PhysRevB.91.214405,PhysRevB.94.134423}, and superfluidity in lattice systems \cite{PhysRevLett.117.045303}, highlighting its role in understanding quantum states in and out of equilibrium.

Meanwhile, quantum quench dynamics induced by a sudden parameter change has been developed into a broad area of study in quantum information, condensed matter physics, and AMO physics \cite{DQPT14,DQPT15,DQPTB2,PhysRevB.93.085416,DQPTreview18,Mitra2018}. By selecting proper quench protocols and initial states, quench systems can exhibit various dynamical phenomena. An example is the dynamical quantum phase transition (DQPT), defined by nonanalytic behavior in real-time dynamics, analogous to the conventional thermodynamic phase transitions \cite{DQPT14,DQPTreview18}. 
Singularities can emerge in the rate function, which plays the role  free energy of thermodynamic systems in dynamical evolution. Theoretical studies and experimental realizations in ionic and atomic systems \cite{DQPTB41,DQPTB4,Zhang_2017,PhysRevApplied.11.044080,PhysRevLett.122.020501,PhysRevB.100.024310,PhysRevLett.124.250601,PhysRevLett.128.160504} have demonstrated DQPTs by revealing quantum orthogonality and phase boundaries with extensions to topological systems \cite{PhysRevB.93.085416,ZianiSR20}. Furthermore, geometric constraints in quench dynamics can induce discontinuous phases under suitable conditions \cite{PhysRevB.110.134319,PhysRevB.111.174310}.

To understand the local geometric evolution in quench dynamics, we study the post-quench QGT of generic one-dimensional (1D) periodic systems and then use the Su-Schrieffer-Heeger (SSH) model~\cite{SSH} as a concrete example.
One interesting feature of the SSH model is that it is a 1D two-band model with a gap-closing point accessible by tuning the ratio of the alternating hopping coefficients. The gap closing point also separate two topologically distinct regimes~\cite{Asboth2016}. There have been theoretical studies of quench dynamics of the SSH model, discussing the fate of localized edge states under open boundary condition ~\cite{Rossi_2022,PhysRevE.108.034102} and emergence of DQPTs~\cite{ZHANG2026131193}, as well as experimental demonstrations of topological solitons through quench dynamics~\cite{Meier2016}. Focusing on the local geometry of quench dynamics, we will show that the QGT of the SSH model can be enhanced when the system is close to the gap closing point, but the post-quench QGT depends also on the parameter change and is further determined by several factors from the quantum states and their associated geometric and thermodynamic quantities.

Since the QGT is invariant under the $U(1)$ gauge transformation from the phase of the quantum state, ordinary unitary evolution generated by the same Hamiltonian would leave the QGT unchanged. However, a quantum quench of the Hamiltonian breaks this symmetry and drives the state along a non-symmetric trajectory on the quantum-state manifold. Consequently, the diagonal component $Q_{kk}$ of a 1D periodic system with crystal momentum $k$ acquires secular terms and gives rise to a metric of the form  $g_{kk}^{(0)}+g_{kk}^{(1)}t+g_{kk}^{(2)}t^{2}$, where the dominant $t^{2}$ term will be shown to be proportional to the group-velocity variance, signaling ballistic wavepacket dispersion. The off-diagonal component $Q_{kt}$ is a purely non-equilibrium quantity featuring a secular linear growth and an oscillatory term with the coefficient associated with the Berry connection, while $Q_{tt}$ coincides with the energy variance and remains independent of time. Our numerical results across various quench protocols confirm that the post-quench QGT is governed by geometric and thermodynamic quantities, such as the Berry connection, group velocity, and energy variance, and reveal their interplays.

The rest of the paper is organized as follows. In Sec.~\ref{Sec:Theory}, we introduce the theoretical framework of the QGT and establish how its components link to operator variances and covariances in the context of quench dynamics. In Sec.~\ref{Sec:SSHQGT}, we apply this formalism to the SSH model, presenting analytical expressions for the post-quench QGT and explaining the physical origin of each term. Numerical results and their physical implications are discussed in Sec.~\ref{Sec:Examples}. We conclude our work in Sec.~\ref{Sec:Conclusion}. The Appendix contains some details and analyses for the dynamics of the QGT.



\section{Theoretical Formalism of the Post-quench QGT}\label{Sec:Theory}
\subsection{Quantum Geometric Tensor}
The quantum geometric tensor (QGT) is a complex second-order tensor that characterizes the geometry of quantum states in parameter space \cite{QGT10}. For a quantum state $|\psi(\boldsymbol{\lambda})\rangle$ parameterized by $\boldsymbol{\lambda}=(\lambda_1,\lambda_2,\cdots)$, the gauge-invariant QGT is defined as
\begin{align}\label{eq:QGT}
Q_{\mu\nu} = \langle \partial_{\mu} \psi | \partial_{\nu} \psi \rangle - \langle \partial_{\mu} \psi | \psi \rangle \langle \psi | \partial_{\nu} \psi \rangle,
\end{align}
with $\partial_{\mu} = \partial / \partial \lambda_{\mu}$. The real part of $Q_{\mu\nu}$, known as the Fubini-Study metric, quantifies the distance between neighboring quantum states, while the imaginary part is proportional to the Berry curvature, capturing topological properties. Specifically, the Fubini-Study metric is given by $g_{\mu\nu} = \text{Re}\,Q_{\mu\nu}$, and the Berry curvature in the $\mu\nu$-space by $\mathcal{F}_{\mu\nu} = -2\,\text{Im}\,Q_{\mu\nu}$. The QGT is invariant under the $U(1)$ gauge transformation $|\psi\rangle \to \me^{\mi \phi} |\psi\rangle$, ensuring physical consistency.

\subsection{Post-quench QGT as Operator Variance and Covariance}\label{Subsec:Operators}
We briefly outline quantum quench dynamics and set $\hbar = 1$ hereafter. A quantum system is initially prepared in $|\psi(0)\rangle$. At $t=0$ the Hamiltonian is suddenly switched, which can be incorporated as $H(t)=H_i\theta(-t)+H_f\theta(t)$, where $\theta(x)$ is the Heaviside step function and $H_{i,f}$ denote the initial and final Hamiltonians, respectively. For $t>0$, the state evolves continuously as $|\psi(t)\rangle = \me^{-\mi H_f t}|\psi(0)\rangle$, satisfying $\mi\partial_t|\psi\rangle=H_f|\psi\rangle$ \cite{MQM}.

When chosen properly, the components of the QGT can be directly linked to physical observables by identifying the parameter derivatives with quantum operators. In this paper we consider a one-dimensional periodic system characterized by crystal momentum $k$ and take $\boldsymbol{\lambda}=(k,t)$ as the parameters for the post-quench QGT.

\subsubsection{Diagonal components $Q_{kk}$ and $Q_{tt}$}
We first consider the $kk$-component of the QGT, which is real and gives the metric $Q_{kk}=g_{kk}$. For the momentum parameter, the derivative $\partial_k$ is related to the position operator \cite{MQM}. For periodic lattice systems, we consider a Bloch state $|\psi_k\rangle = \sum_n \me^{\mi k n}|n\rangle \otimes |u_k\rangle$, 
where $|n\rangle$ labels the unit cells and $|u_k\rangle$ is the cell-periodic spinor satisfying $u_k(x+a) = u_k(x)$ with lattice constant $a$. The maximally localized Wannier function centered at one selected cell is $|w_0\rangle = N^{-1/2}\sum_k \me^{-\mi k \hat{x}}|\psi_k\rangle$, 
where $\hat{x}$ is the position operator. Standard results \cite{Blount1962,Resta1994,Vanderbilt2018} show that the expectation value of $\hat{x}$ in this Wannier state gives the Berry connection
$\langle w_0|\hat{x}|w_0\rangle = \mathrm{i}\langle \partial_k u_k|u_k\rangle \equiv \mathcal{A}(k)$
while the variance of $\hat{x}$ equals the quantum metric:
\begin{align}\label{gkk0}
\operatorname{Var}(\hat{x})
&=\langle w_0|\hat{x}^2|w_0\rangle - \langle w_0|\hat{x}|w_0\rangle^2 \notag\\
&= \langle \partial_k u_k|\big(1-|u_k\rangle\langle u_k|\big)|\partial_k u_k\rangle \notag \\
&=g_{kk}.
\end{align}
Thus, in momentum space the operator $\mathrm{i}\partial_k$ acts as the position operator $\hat{x}$ for the wave-packet center of mass, and $g_{kk}$ measures the variance of the center-of-mass position.

During the post-quench evolution, the state becomes $|\psi_k(t)\rangle=\me^{-\mi H_f t}|\psi_k\rangle$. 
In the Heisenberg picture, the time evolution of the position operator is governed by the Heisenberg equation of motion, $\frac{\dif\hat{x}(t)}{\dif t} = \mi [H_f, \hat{x}(t)] = \hat{v}_k$. Here we introduce the group-velocity operator
$\hat v_k \equiv \partial_k H_f(k)$,
which acts on the Hilbert space at fixed $k$.
Since $\hat{v}_k$ does not explicitly depend on time for a static $H_f$, this equation integrates directly to
\begin{align}\label{eq:xHeis}
\hat{x}(t)=\hat{x}+\hat{v}_k t.
\end{align}
The time-dependent quantum metric $g_{kk}(t)$ is the variance of this evolved position operator, generalizing the static definition in Eq.~\eqref{gkk0}. It therefore expands as
\begin{align}\label{Eq:gkkt}
g_{kk}(t)
&= \operatorname{Var}(\hat{x}) + 2t\,\operatorname{Cov}(\hat{x},\hat{v}_k)
 + t^2\,\operatorname{Var}(\hat{v}_k) \notag\\
&\equiv g_{kk}^{(0)}(k,t) + g_{kk}^{(1)}(k,t)\,t + g_{kk}^{(2)}(k)\,t^2.
\end{align}
Here the expectation values such as $\langle \hat{v}_k\rangle$ and $\langle \hat{v}_k^2\rangle$ are taken with respect to $|\psi_k(t)\rangle$. 
The initial geometric uncertainty $g_{kk}^{(0)}=\operatorname{Var}(\hat{x})$ is dragged by the group-velocity field, leading to a ballistic $t^2$ growth that dominates at long times. This operator-variance interpretation unifies abstract quantum geometry with the tangible picture of wave-packet spreading in real space.
We mention that the Fisher information, which plays a similar role as the metric from the QGT, has been shown to have a $t^2$ dependence with the coefficient as a variance~\cite{PhysRevLett.124.060402}.

In the absence of a quench, the time evolution $\exp(-\mi H_i t)$ merely multiplies each momentum eigenstate by a phase factor, i.e., it is a $U(1)$ gauge transformation along the $k$-direction.
Consequently, $g_{kk}$ remains invariant along this evolution, meaning the state moves along a symmetry-preserving orbit on which the distance between neighboring momentum states is constant with respect to $t$.
After a quench, however, the evolution operator $U(t)=\exp(-\mi H_f t)$ generates a unitary rotation in the two-dimensional Hilbert space spanned by the post-quench ground and excited states at each momentum $k$. Moreover, the initial state is decomposed into the two eigenstates of $H_f$, making the post-quench time-volution operator effectively a $U(2)$ transformation.
Since $[H_i, H_f]\neq0$ for nontrivial cases, the time evolution generated by $h_f$ is no longer a symmetry of the initial Hamiltonian. Hence, the system is driven away from the original symmetric orbit after the quench.
Such a quench-induced geometric deviation is quantified by the linear term $g_{kk}^{(1)}\propto\operatorname{Cov}(\hat{x},\hat{v}_k)$, as well as the quadratic term $g_{kk}^{(2)}=\operatorname{Var}(\hat{v}_k)$. Thus, the post-quench dynamics follows a non-symmetric trajectory (with respect to $H_i$) on the quantum state manifold, along which the local distance between different momentum states varies with time.

The temporal component $Q_{tt}$ is also real and defines the metric $g_{tt}$. A straightforward evaluation using the QGT definition, Eq.~\eqref{eq:QGT}, shows that it has the simple form
\begin{equation}\label{Eq:qtt}
g_{tt} = \operatorname{Var}(H_f)=\Delta E^2,
\end{equation}
where $\Delta E^2 \equiv \langle H_f^2\rangle - \langle H_f\rangle^2$ is the variance of the post-quench Hamiltonian with respect to the time-evolved state, which can be shown to be the same as that with respect to the initial state by an expansion in the post-quench eigenstates. Thus, $g_{tt}$ is time-independent and directly encodes the energy uncertainty introduced by the quench.

\subsubsection{Off-diagonal component $Q_{kt}$}
The off-diagonal component $Q_{kt}$ similarly admits an operator representation. Substituting the post-quench Schr\"odinger equation, $\partial_t|\psi\rangle = -\mathrm{i}H_f|\psi\rangle$, into the definition of the QGT yields $Q_{kt} = -\mathrm{i} \bigl( \langle \partial_k \psi | H_f | \psi \rangle - \langle \partial_k \psi | \psi \rangle \langle H_f \rangle \bigr)$. 
Recognizing the position operator $\hat{x} = \mathrm{i}\partial_k$ (so that $\langle \partial_k \psi | = \mathrm{i}\langle \psi | \hat{x}$), this simplifies to the covariance between $\hat{x}$ and $H_f$:
\begin{align}
Q_{kt} = \langle \hat{x} H_f \rangle - \langle \hat{x} \rangle \langle H_f \rangle = \operatorname{Cov}(\hat{x}, H_f).
\end{align}
This covariance can be decomposed into symmetric and antisymmetric parts
$Q_{kt} = 
\tfrac{1}{2}\langle \{\hat{x}, H_f\} \rangle - \langle \hat{x} \rangle \langle H_f \rangle
\;+\; \tfrac{1}{2}\langle [\hat{x}, H_f] \rangle$.
The first term is the symmetrized covariance while the second proportional to $\langle [\hat{x}, H_f] \rangle = \mathrm{i} \langle \hat{v}_k \rangle$ stems from the non-commutativity of position and the Hamiltonian. Through the connections to variance and covariance, the post-quench QGT is endowed with physical meanings and implications.

The operator formalism leads to a fundamental inequality unifying all three components of the QGT:
\begin{align}
Q_{kk} Q_{tt} \ge |Q_{kt}|^{2}.
\end{align}
Since $Q_{tt}= \operatorname{Var}(H_f)$, $g_{kk}= \operatorname{Var}(\hat{x})$, and $Q_{kt}= \operatorname{Cov}(\hat{x}, H_f)$, this inequality reflects the semi-positivity of the quantum variance-covariance matrix:
$\operatorname{Var}(\hat{x}) \cdot \operatorname{Var}(H_f) \ge |\operatorname{Cov}(\hat{x}, H_f)|^2$.
Physically, the inequality bounds the values of the variance and covariance in the post-quench dynamics. 
The constraint also highlights the subtle relation between geometric spreading and energy variation through the covariance. 
We mention that Ref.~\cite{PhysRevB.104.045103} presents other inequalities involving the QGT and Chern number for higher-dimensional systems.

\section{QGT dynamics of the SSH Model}\label{Sec:SSHQGT}

\subsection{The SSH Model}
We now apply the general formalism to the 1D two-band Su-Schrieffer-Heeger (SSH) model with periodic boundary condition. The Bloch Hamiltonian is $H(k) = \mathbf{R}(k) \cdot \boldsymbol{\sigma}$, where $\mathbf{R}(k) = (-J_1 - J_2 \cos k,\ J_2 \sin k,\ 0)^\mathrm{T}$. Introducing the dimerization parameter $m = J_1/J_2$ and $\tilde{R} = \sqrt{m^2 + 1 + 2m \cos k}$, the eigenvalues are $E_{\pm} = \pm J_2 \tilde{R}$ with corresponding eigenstates
\begin{align}
|u^{\pm}_k\rangle = \frac{1}{\sqrt{2} \tilde{R}} \begin{pmatrix} \tilde{R} \\ \mp(m + \me^{-\mi k}) \end{pmatrix}.
\end{align}
Here, $|u^-_k\rangle$ ($|u^+_k\rangle$) denotes the ground (excited) state. The point $m = 1$ marks the gap closing. For $m < 1$ ($m > 1$), the system is topologically nontrivial (trivial) with a quantized winding number \cite{Asboth2016}.

We study the dynamics following a sudden quench of $m$: the system is prepared in the ground state $|u^{i,-}_k\rangle$ of the initial Hamiltonian $H_i$ with parameter $m_i$. At $t=0$, the parameter is abruptly changed to $m_f$ (keeping $J_2>0$ constant), and subsequent evolution is governed by the final Hamiltonian $H_f$. The eigenstates of $H_f$, $|u^{f,\pm}_k\rangle$, have energies $\mp R_f = \mp J_2 \tilde{R}_f$, where $\tilde{R}_f = \sqrt{1 + m_f^2 + 2 m_f \cos k}$.

The time-evolved state is $|\psi_k(t)\rangle = \me^{-\mi H_f t} |u^{i,-}_k\rangle$. Following Ref.~\cite{DTQPT18} for handling quench dynamics of periodic systems, we decompose the initial state in the final basis via $|u^{i,-}_k\rangle = \alpha |u^{f,-}_k\rangle + \beta |u^{f,+}_k\rangle$  with the coefficients
\begin{align}
\alpha, \beta = \frac{1}{2 \tilde{R}_i \tilde{R}_f} \left[ \tilde{R}_i \tilde{R}_f \pm (m_i + \me^{-\mi k})(m_f + \me^{\mi k}) \right],
\end{align}
where $\tilde{R}_i = \sqrt{m_i^2 + 1 + 2 m_i \cos k}$, and the $+$ ($-$) sign corresponds to $\alpha$ ($\beta$). Although $\alpha$ and $\beta$ appear singular when $m_{i,f}=1$ and $k=\pi$, both numerator and denominator vanish simultaneously, yielding finite values upon careful limiting. Thus,
\begin{align}\label{e13}
|\psi_k(t)\rangle = \alpha \me^{+\mi R_f t} |u^{f,-}_k\rangle + \beta \me^{-\mi R_f t} |u^{f,+}_k\rangle.
\end{align}
We compute the QGT using $(k, t)$ as parameters in Eq.~\eqref{eq:QGT}. In the post-quench eigenbasis, $v_k^\pm = \pm \partial_k R_f(k)$ represent the group velocities. Moreover, $\langle \hat{v}_k \rangle = |\alpha|^2 v_k^- + |\beta|^2 v_k^+$. 

\subsection{Diagonal component $Q_{kk}$}
For the SSH model, applying $\mathrm{i}\partial_k$ to Eq.~(\ref{e13}) gives two contributions:
\begin{align}
\mathrm{i}\partial_k|\psi_k(t)\rangle = &\alpha \me^{\mathrm{i}R_f t}\bigl(\mathrm{i}\partial_k|u^{f,-}_k\rangle\bigr) + \beta \me^{-\mathrm{i}R_f t}\bigl(\mathrm{i}\partial_k|u^{f,+}_k\rangle\bigr) \notag \\
+& t\bigl(\partial_k R_f\bigr)\bigl(\alpha \me^{\mathrm{i}R_f t}|u^{f,-}_k\rangle - \beta \me^{-\mathrm{i}R_f t}|u^{f,+}_k\rangle\bigr).\notag
\end{align}
The first line contains the geometric part, involving the Berry connections of the post-quench eigenstates. The second line is proportional to the group-velocity operator, since $\partial_k R_f$ ($-\partial_k R_f$) represents the group velocity of the excited (ground) band after the quench. Denoting the geometric part of the first line by $(\mathrm{i}\partial_k)_{\text{geom}}$, we have
\begin{align}
\mathrm{i}\partial_k|\psi_k(t)\rangle = (\mathrm{i}\partial_k)_{\text{geom}}|\psi_k(t)\rangle + t\,\hat{v}_k|\psi_k(t)\rangle.
\end{align}
Using Eq.~\eqref{eq:QGT}, the quantum metric therefore expands as
\begin{align}\label{23}
g_{kk}(t)
= &\bigl\langle(\mathrm{i}\partial_k)_{\text{geom}}^2\bigr\rangle - \bigl\langle(\mathrm{i}\partial_k)_{\text{geom}}\bigr\rangle^2 \notag \\
+& 2t\Bigl(\bigl\langle(\mathrm{i}\partial_k)_{\text{geom}} v_k\bigr\rangle - \bigl\langle(\mathrm{i}\partial_k)_{\text{geom}}\bigr\rangle\langle \hat{v}_k\rangle\Bigr) \notag \\
+& t^2\Bigl(\langle \hat{v}_k^2\rangle - \langle \hat{v}_k\rangle^2\Bigr).
\end{align}
A direct evaluation confirms the form 
of Eq.~(\ref{Eq:gkkt}) with each coefficient containing oscillatory terms that encode coherent oscillations between the post-quench bands.

The zeroth-order term in $t$ of $g_{kk}$ is
\begin{align}
g_{kk}^{(0)}(k,t) = & g_{kk}^i(k) + 4 \left( \mathcal{B}^2 - \mathcal{A}_i \mathcal{A}_f \right) \sin^2(R_f t) \notag \\
& + 4 \left( \mathcal{A}_f^2 - \mathcal{B}^2 \right) \sin^4(R_f t),
\end{align}
where $g_{kk}^i(k) = \frac{(m_i \cos k + 1)^2}{4 \tilde{R}_i^4}$ is the pre-quench metric,
\begin{align}
\mathcal{A}_\lambda = \mi \langle u^{\lambda,-}_k | \partial_k u^{\lambda,-}_k \rangle = \frac{m_\lambda \cos k + 1}{2 \tilde{R}_\lambda^2}
\end{align}
(with $\lambda = i, f$) is the Berry connection of the pre- ($i$) or post-quench ($f$) ground state, and
\begin{align}
\mathcal{B} = &-\mathcal{A}_f \frac{(m_i m_f + m_i \cos k + m_f \cos k + 1)}{\tilde{R}_i \tilde{R}_f}\notag\\ =& -\mathcal{A}_f (2 \operatorname{Re}[\alpha] - 1).
\end{align}
The term $g_{kk}^{(0)}$ encapsulates the instantaneous geometric response, featuring bounded, non-secular oscillations at frequency $R_f$ due to coherence between the post-quench states. These oscillations, modulated by $\mathcal{A}_i$, $\mathcal{A}_f$, and $\mathcal{B}$, signify interference between initial and final state geometries. We note that the pre-quench metric $g_{kk}^i$ characterizes the width of the initial wave packet at momentum $k$ and is enhanced near the gap-closing point.

The coefficient of the first-order term is
\begin{align}\label{Eq:gkk1}
g_{kk}^{(1)}(k,t) = 2 \mathcal{B} \mathcal{C} \sin(2 R_f t),
\end{align}
with
\begin{align}
\mathcal{C} = \frac{J_2 m_f (m_i - m_f) \sin^2 k}{\tilde{R}_i \tilde{R}_f^2} = 2 v_k^+ |\alpha \beta| \operatorname{sign}(m_i - m_f),
\end{align}
where $v_k^+ = \partial_k R_f = \frac{J_2 m_f \sin k}{\tilde{R}_f}$ represents the group velocity of the post-quench excited band. Equation \eqref{Eq:gkk1} signifies a linear-in-time modulation arising from the interplay between group velocity (through $\mathcal{C}$) and Berry connection (through $\mathcal{B}$), with the $\sin(2 R_f t)$ factor reflecting sustained quantum coherence.

For the SSH model, the coefficient $g_{kk}^{(2)}(k)
= \operatorname{Var}(\hat{v}_k)$ simplifies to
\begin{align}
g_{kk}^{(2)}(k)
=& 4 J_2^2 \frac{m_f^2 \sin^2 k}{\tilde{R}_f^2} |\alpha \beta|^2 \notag \\
=& |\alpha|^2 (v_k^-)^2 + |\beta|^2 (v_k^+)^2 - \left( |\alpha|^2 v_k^- + |\beta|^2 v_k^+ \right)^2 \notag \\
=& \mathcal{C}^2.
\end{align}
This variance contains the quantum uncertainty factor $|\alpha \beta|^2$ and characterizes the fluctuations induced by the non-equilibrium superposition. It drives a secular $t^2$ growth due to the velocity-velocity correlations, leading to ballistic separation of the ground- and excited- band components in momentum space.

The expansion of $g_{kk}(t)$ offers a physical interpretation unified with the operator picture of Sec.~\ref{Subsec:Operators}. The operator $\mathrm{i}\partial_k$ represents the center-of-mass position of the wavepacket, making $g_{kk}$ the total variance. The zeroth-order term $g_{kk}^{(0)}$ combines the initial positional variance ($g_{kk}^i$) with bounded oscillations from inter-band coherence. The coefficient of the first-order term $g_{kk}^{(1)}$ is proportional to the covariance $\operatorname{Cov}((\mathrm{i}\partial_k)_{\text{geom}}, \hat{v}_k)$, oscillating with frequency $2R_f$ due to the transition between the post-quench bands. Finally, the second-order term $g_{kk}^{(2)}t^2 = \operatorname{Var}(\hat{v}_k)t^2$ dominates at long times, capturing the ballistic wavepacket dispersion that manifests as real-space motion.
The structure of $g_{kk}$ reveals the interplay between quantum geometry and nonequilibrium dynamics: The oscillatory terms capture the coherent response, while the secular (linear and quadratic) terms describe the progressive evolution. 

We remark that without a quench ($m_i=m_f$), one finds $\mathcal{A}_i = \mathcal{A}_f = -\mathcal{B}$ and $\mathcal{C}=0$, causing $g_{kk}^{(1)}$ and $g_{kk}^{(2)}$ to vanish. The metric then reduces to the static pre-quench value $g_{kk}^i$, consistent with unitary evolution under the same Hamiltonian preserving the quantum geometry at each $k$. A general proof is provided in Appendix \ref{app2}.

\subsection{Diagonal component $Q_{tt}$}
Using the pre-quench ground state $|u^{i,-}_k\rangle$,
the $tt$-component of the post-quench QGT for the SSH model can be obtained by Eq.~\eqref{Eq:qtt}. Explicitly,
it is
\begin{align}
g_{tt} = 4 J_2^2 \tilde{R}_f^2 |\alpha|^2 |\beta|^2  = J_2^2 \frac{(m_i - m_f)^2 \sin^2 k}{\tilde{R}_i^2}.
\end{align}


In the absence of a quench ($m_i=m_f$), it follows that $Q_{tt}=0$, recovering the static limit. Moreover, $Q_{tt}$ diverges at the gap closing point ($m_i=1$), reflecting the enhancement of energy fluctuations when the bands touch each other.

\subsection{Off-diagonal component $Q_{kt}$}\label{Sec:SSHQkt}
The off-diagonal component $Q_{kt}$ characterizes the geometric cross-response between momentum and time. For the SSH model, its explicit form is given by
\begin{align}
\operatorname{Re} Q_{kt} &= \mathcal{D} \big[ \mathcal{B} \sin(2 R_f t) + \mathcal{C} t \big], \label{eq:ReQkt} \\
\operatorname{Im} Q_{kt} &= \mathcal{D} \big( \mathcal{A}_i - 2 \mathcal{A}_f \sin^2 (R_f t) \big). \label{eq:ImQkt}
\end{align}
Here, the quench-activated prefactor is
\begin{align}\label{eq:D}
\mathcal{D} = -\sqrt{\Delta E^2} \cdot \operatorname{sign}\!\big((m_i - m_f) \sin k\big),
\end{align}
which is odd in momentum and vanishes in the absence of a quench ($m_i = m_f$).

These expressions reveal the distinct physical information encoded in $Q_{kt}$. The real part combines an oscillatory term, mediated by the post-quench Berry connection $\mathcal{B}$, with a secular linear growth term proportional to $\mathcal{C}$. Since $\mathcal{C}$ is related to $\sqrt{\operatorname{Var}(\hat{v}_k)}$, this linear term directly tracks the velocity fluctuations of the wavepacket and the associated dispersion in momentum space. The imaginary part corresponds to a time-dependent Berry curvature $\mathcal{F}_{kt} = -2\operatorname{Im} Q_{kt}$ on the extended $(k,t)$ parameter space. Its structure combines a constant offset from the initial geometry ($\propto \mathcal{A}_i$) with coherent oscillations driven by the final Hamiltonian ($\propto \mathcal{A}_f$). The time average $\langle \operatorname{Im} Q_{kt} \rangle = \mathcal{D}(\mathcal{A}_i - \mathcal{A}_f)$ thus quantifies the net geometric shift induced by the quench.

For the SSH model, $Q_{kt}$ is an odd function with respect to $k$ because of the prefactor $\mathcal{D}$ given in Eq.~\eqref{eq:D}. The factor $\operatorname{sign}(\sin k)$ within $\mathcal{D}$ is odd under $k \to -k$. This specific form stems from the analytical derivation of $Q_{kt}$ for the SSH model. The other constituents of $Q_{kt}$, namely the Berry connections $\mathcal{A}_i(k)$, $\mathcal{A}_f(k)$, and the coefficients $\mathcal{B}(k)$, $\mathcal{C}(k)$, are even functions of $k$. This is evident from their dependence on $\cos k$ and $\tilde{R}_\lambda^2 = m_\lambda^2 + 1 + 2 m_\lambda \cos k$ for $\lambda=i,f$ in the expressions shown in Sec.~\ref{Sec:SSHQGT}. Moreover, given that $\operatorname{Im} Q_{kt}(k, t)$ 
for the SSH model is an odd function of $k$, its integral over the Brillouin zone must vanish. This confirms the absence of net geometric phase accumulation from the quench-induced curvature. It is crucial to distinguish the curvature $\mathcal{F}_{kt} = -2\operatorname{Im} Q_{kt}$ generated by quench dynamics from the adiabatic Berry phase of the SSH model. The latter is a global topological invariant ($\pi$ for $m<1$, $0$ for $m>1$), whereas $\mathcal{F}_{kt}$ is a local curvature in the extended $(k,t)$ space, induced by the quench. Even when the adiabatic Berry phase is zero (e.g., for $m>1$), the quench generates nonzero local curvature fluctuations which integrate to zero due to the odd symmetry of the curvature with respect to $k$.
The dynamics of $Q_{kt}$ thus encapsulate the interplay between memory of the initial geometry (through $\mathcal{A}_i$), sustained  coherence (through the oscillatory terms in $\mathcal{B}$ and $\mathcal{A}_f$), and ballistic spreading (through the linear term in $\mathcal{C}$). 

In summary, the post-quench QGT provides a comprehensive geometric representation of suitably chosen physical quantities. Its components reflect the energy variance ($Q_{tt}$), wavepacket dispersion ($Q_{kk}$), and momentum-time correlations ($Q_{kt}$), highlighting how the quantum geometry of a nonequilibrium state encodes the uncertainties and cross-correlations of its underlying observables.

\section{Numerical examples}\label{Sec:Examples}
Here we show exemplary post-quench QGT results for the SSH model when the parameter $m$ is suddenly changed.

\begin{figure*}[t]
\centering
\includegraphics[width=3.5in, clip]{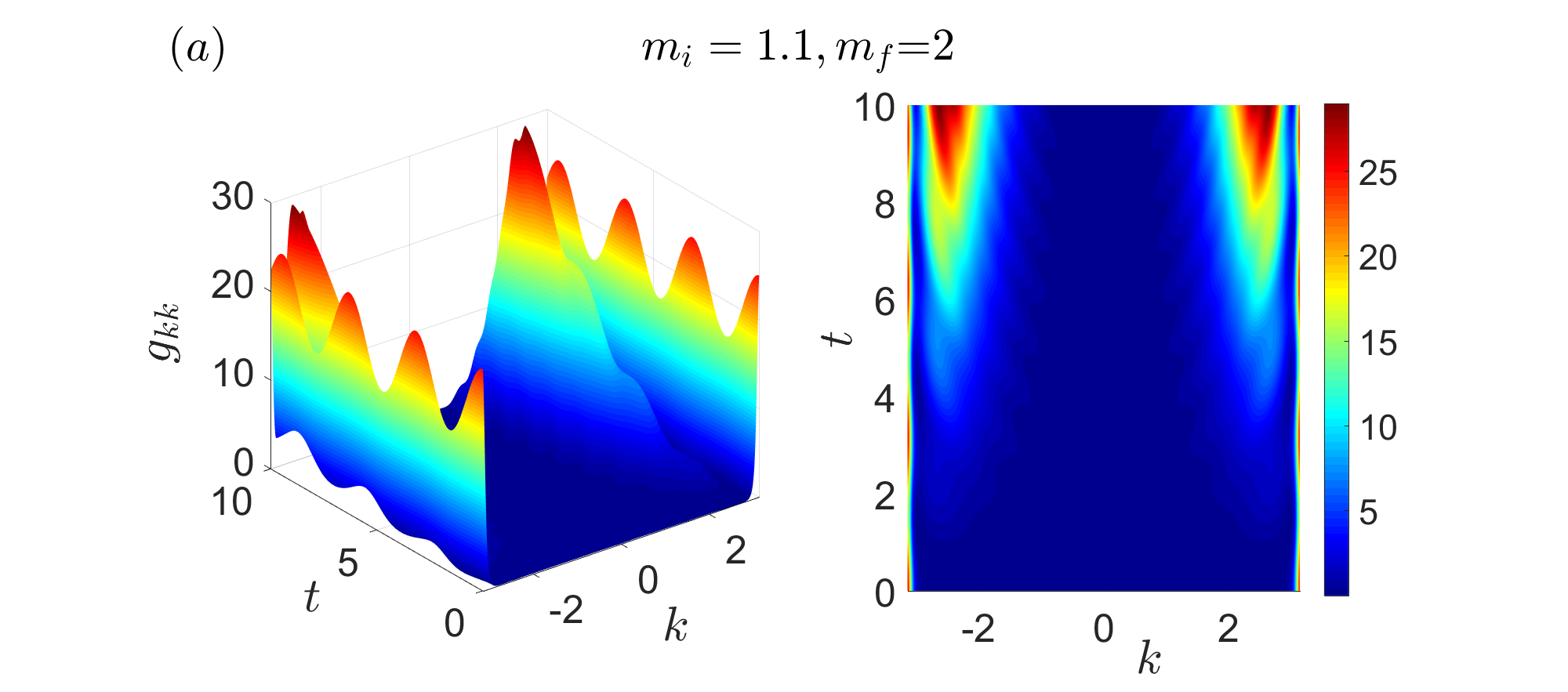}
\includegraphics[width=3.5in, clip]{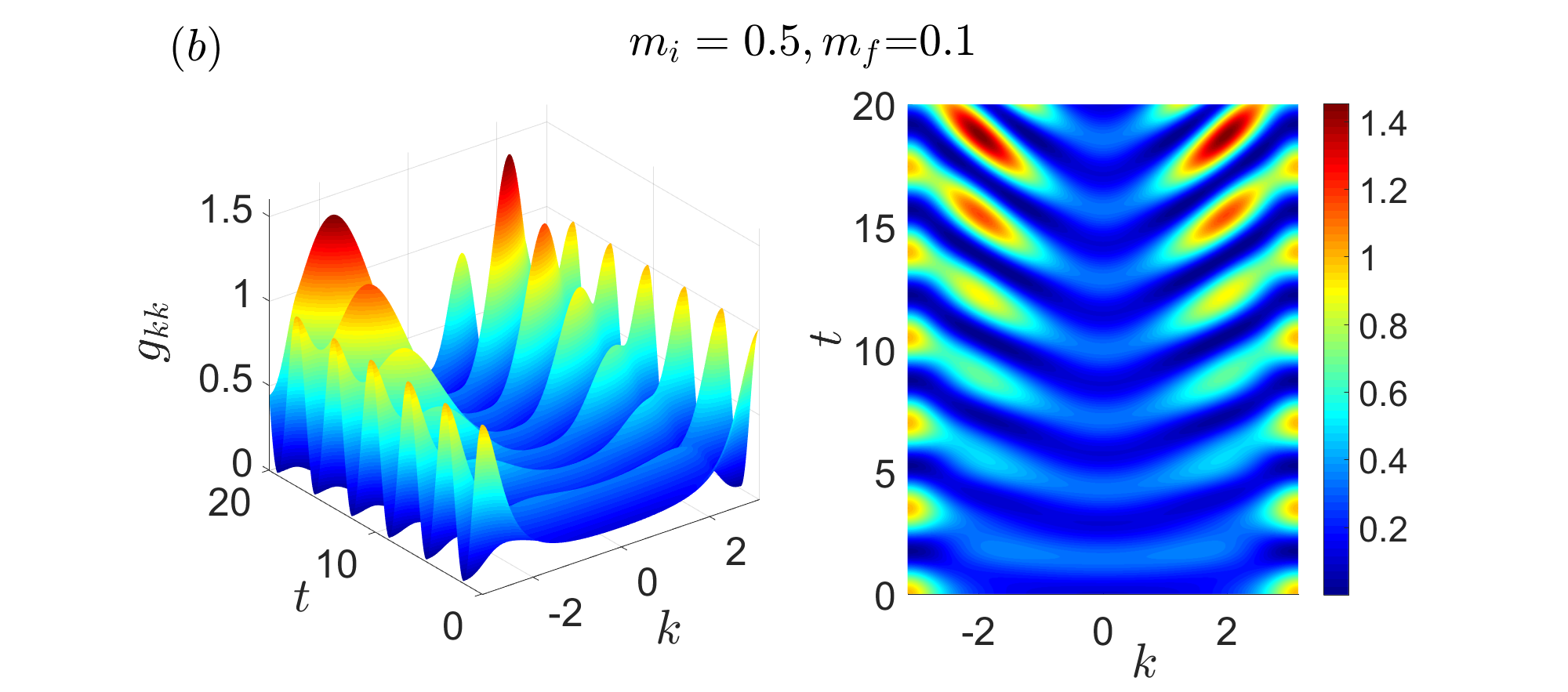}\\
\includegraphics[width=3.5in, clip]{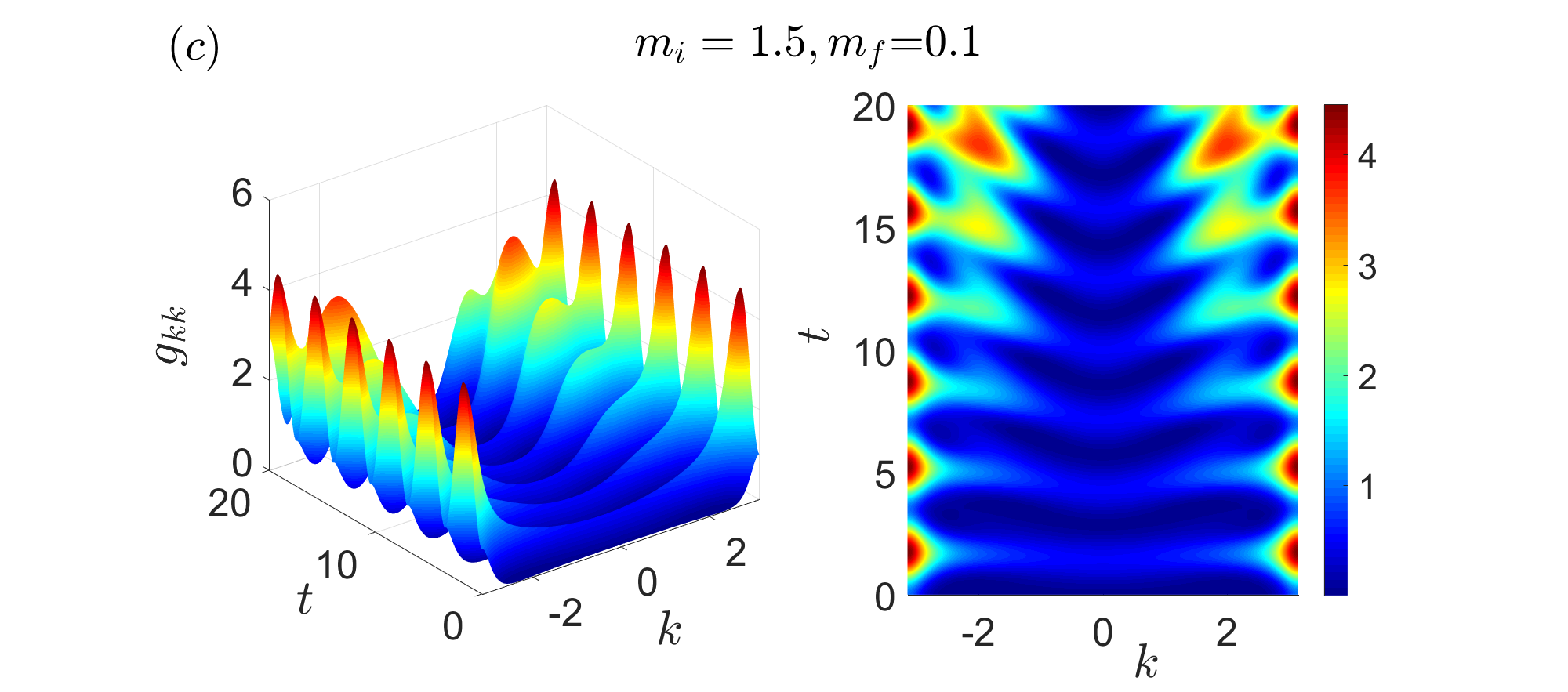}
\includegraphics[width=3.5in, clip]{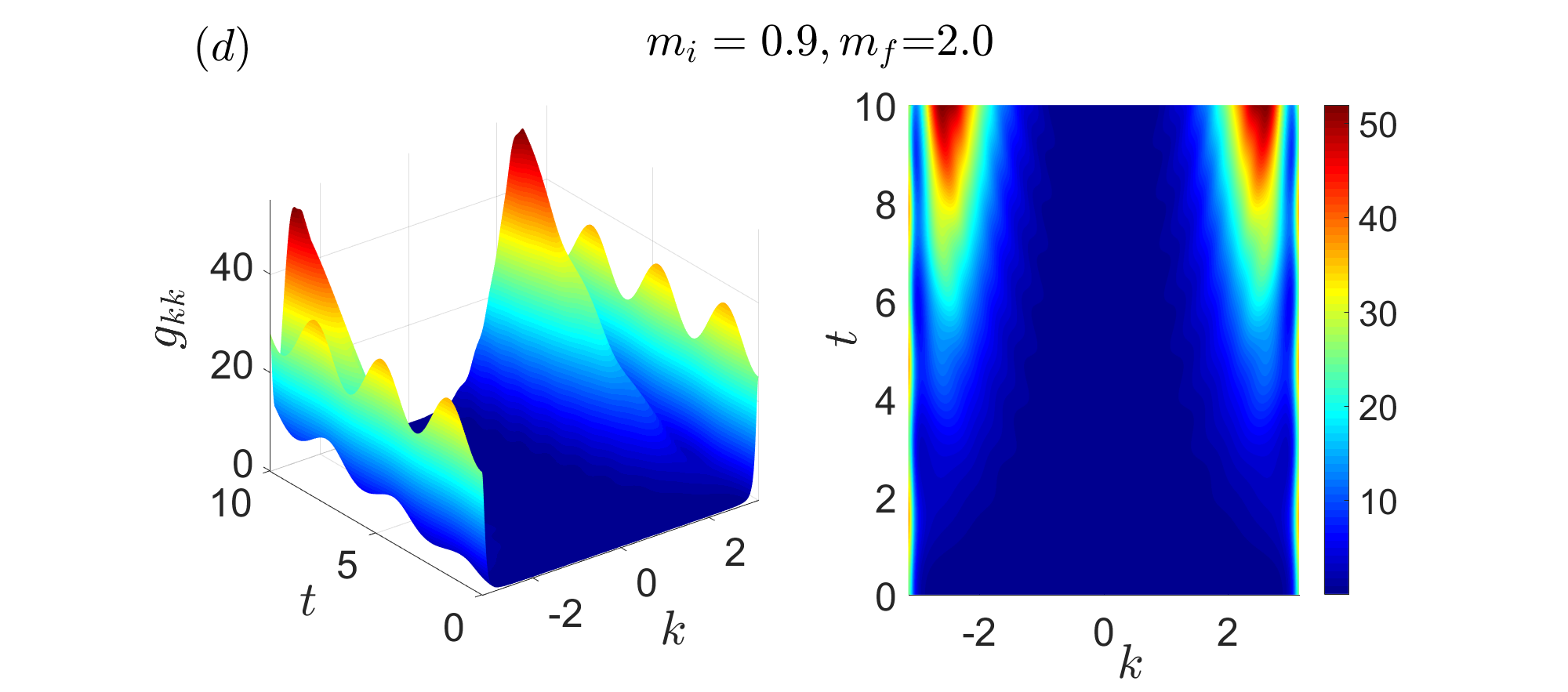}
\caption{Quantum-metric dynamics $g_{kk}(k,t)$ for four quench protocols.
For each case, the left (right) panel shows the 3D surface (contour map).
The top (bottom) row shows quenches staying on the same side of (crossing) the gap closing point $m=1$. The values of $m_i$ and $m_f$ are labeled above each case.
}
\label{Fig_g_kk}
\end{figure*}

\subsection{Diagonal component $Q_{kk}$}
In Fig.~\ref{Fig_g_kk} we plot the behavior of $g_{kk}$ in the $(k,t)$-plane for four distinct quench protocols. In each sub-panel, the left-hand side gives a three-dimensional view and the right-hand side shows the corresponding contour map. The two upper panels correspond to quenches between phases with identical topological character, whereas the two lower panels correspond to quenches across the topological phase boundary. The four panels together illustrate how the relative weight of the oscillatory terms $g_{kk}^{(0)},g_{kk}^{(1)}t$ and the ballistic term $g_{kk}^{(2)}t^{2}$ depends on the quench protocol. Recall that the quadratic coefficient $g_{kk}^{(2)}=\mathcal{C}^{2}= \operatorname{Var}(\hat{v}_{k})$ is independent of time; it is fixed by the post-quench bands and the initial state. When $g_{kk}^{(2)}$ is large, the $t^{2}$ growth quickly dominates the short-time dynamics and overwhelms the oscillatory contributions, producing a smooth, steep ridge. When $g_{kk}^{(2)}$ is modest, the oscillations remain visible and modulate the ridge.

Fig.~\ref{Fig_g_kk} (a) ($m_i=1.1\rightarrow m_f=2.0$) shows the post-quench $g_{kk}$ for a quench above the gap-closing point. Near the Brillouin zone boundaries $k=\pm\pi$, sharp, narrow peaks oscillate periodically in time, stemming solely from the zeroth-order term $g_{kk}^{(0)}$ since $\sin k=0$ extinguishes $\mathcal{C}$ and suppresses secular growth. Away from the boundary, around $|k|\approx 2$, two symmetric ridges grow quadratically with time. Because the quench starts near the critical point with a large parameter jump, the group-velocity variance $\operatorname{Var}(\hat{v}{k})$ is very large, causing the $g{kk}^{(2)}t^{2}$ term to overwhelm the oscillatory contributions on the plotted time scale ($t\le20$). Consequently, the ridges rise steeply and appear almost smooth.

Fig.~\ref{Fig_g_kk} (b) ($m_i=0.5\rightarrow m_f=0.1$) corresponds to a quench well below the gap-closing point. Here the boundary peaks at $k=\pm\pi$ are much lower and broader, consistent with a pre-quench Hamiltonian far from criticality. The dynamics remain governed by coherent oscillations without secular growth at these momenta. In the intermediate region $|k|\approx 2$, the ridge grows quadratically but is strongly modulated by visible oscillations, creating a rippled texture along the time axis. 

Fig.~\ref{Fig_g_kk}(b) ($m_i=0.5\to m_f=0.1$) and Fig.~\ref{Fig_g_kk}(c) ($m_i=1.5\to m_f=0.1$) both exhibit modulated ridges, reflecting a regime where the group-velocity variance $\operatorname{Var}(\hat{v}_k)=g_{kk}^{(2)}$ is only modest. In (b) the boundary peaks are low and broad, consistent with an initial state far from the gap-closing point; in (c) they are moderate and broad because $m_i=1.5$ lies well above the gap closing point. In both cases the oscillatory terms $g_{kk}^{(0)}$ and $g_{kk}^{(1)}t$ remain comparable to $g_{kk}^{(2)}t^2$ over the plotted time window, imprinting visible ripples on the quadratically growing ridge. The resulting competition between momentum-space dispersion and persistent inter-band oscillations confirms that the response is governed by geometric and energetic quantities, not by the change of a topological invariant across the quench.

Fig.~\ref{Fig_g_kk}(d) depicts a distinct scenario of a quench from an initial state close but below the gap closing point to a final state above the gap closing point ($m_i=0.9 \rightarrow m_f=2.0$). This protocol involves a larger parameter change across $m=1$, exhibits the most extreme dynamical response in our selected cases. At $k=\pm\pi$, the oscillatory peaks are exceptionally high and narrow, a direct manifestation of the nearly-divergent pre-quench metric $g_{kk}^i$ when $m_i \approx 1$. Concurrently, the ridges near $|k|\approx 2$ exhibit the fastest quadratic growth among all cases. Similar to panel (a), the ridges are narrow in momentum and smooth because the relatively large $\operatorname{Var}(\hat{v}_{k})$ once again ensures the $t^2$ growth dominates the oscillatory terms. The similarity with case (a) lies in the proximity of the initial state to the gap closing point ($m_i \approx 1$), which dictates the large group-velocity variance and hence the smooth, ballistic-like ridges. However, a difference from the specific geometric structures of the initial states ($m_i < 1$ versus $m_i > 1$) enhances the boundary oscillations via $g{kk}^i$ in (d) and makes the overall response more pronounced than (a).

The ballistic $ t^2 $ growth of $g_{kk}$ in these cases implies that the quench drives the state away from the U(1) orbit of $H_i$ as the wavepacket spreads in momentum and real space. In contrast, without a quench the evolution remains on the symmetry-preserving orbit and $g_{kk}$ stays constant. The secular growth of the quantum metric therefore reflects the progressive separation induced by the velocity mismatch due to the post-quench Hamiltonian and offers a geometric signature of the unitary evolution.

The smooth or rippled ridge in the momentum-resolved evolution of $g_{kk}$ mirrors the magnitude of the time-independent coefficient $g_{kk}^{(2)}=\operatorname{Var}(\hat{v}_k)$. Larger group-velocity variance in Fig.~\ref{Fig_g_kk} (a) and (d) suppresses the oscillations and yields a smooth, ballistic spreading while smaller variance of the cases shown in panels (b) and (c) allows the coherent oscillations to persist, producing a modulated ridge. The similarities between Fig.~\ref{Fig_g_kk}(a) and (d) and those between (b) and (c) show that the decisive factors are the initial proximity to the gap-closing point and the size of the parameter change across the quench, not simply the crossing of the topological phase boundary of the Hamiltonian.

\begin{figure}[t]
\centering
\includegraphics[width=\columnwidth]{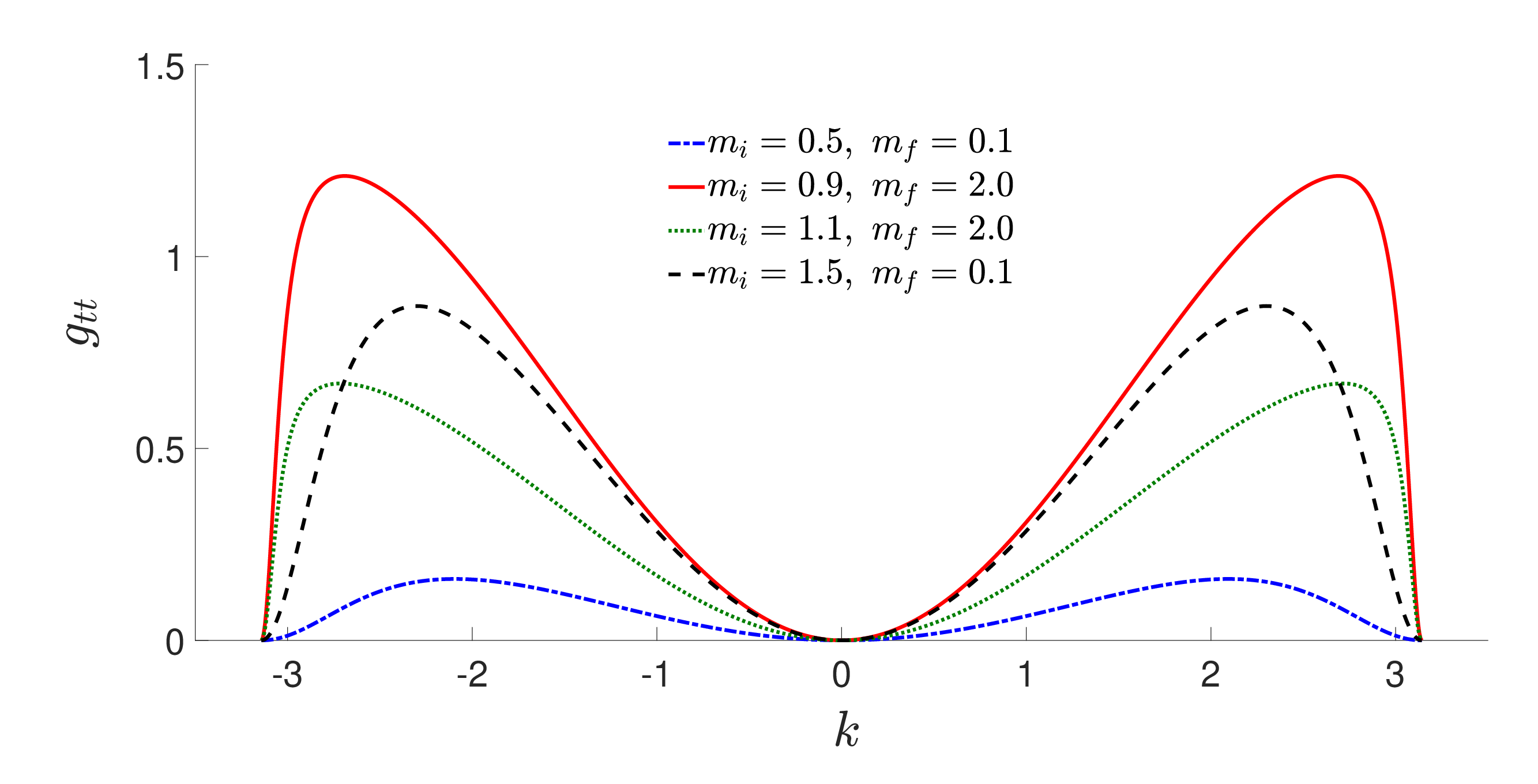}
\caption{Post-quench temporal metric $g_{tt}(k)$.
Blue dot-dash line: $m_i=0.5, m_f=0.1$; Green dotted line: $m_i=1.1, m_f=2.0$; Black dash line: $m_i=1.5, m_f=0.1$; Red solid line: $m_i=0.9, m_f=2.0$.
}
\label{Fig_gtt}
\end{figure}

\subsection{Diagonal component $Q_{tt}$}\label{Sec4B}
The post-quench temporal metric $g_{tt}(k)$ of the SSH model, shown in Fig.~\ref{Fig_gtt}, is symmetric about $k=0$, vanishes at the high-symmetry points $k = 0,\pm\pi$, and displays a single positive peak in each momentum half-space. While $g_{tt}(k) = \Delta E^2(k)$ is constant in time, $\Delta E(k)$ introduces a timescale $\tau_k \sim 1/\Delta E(k)$ for each momentum mode. Such a time scale stems from a fundamental quantum limit known as the Mandelstam-Tamm inequality~\cite{MandelstamTamm45}, which bounds the minimal time for a state to undergo a significant change by $\sim \hbar / \Delta E$. Thus, a larger $g_{tt}(k)$ implies a shorter $\tau_k$, marking the corresponding mode more volatile. It is crucial to note that this $ \tau_k $ quantifies an upper bound for a significant change, whereas the actual oscillation frequency in the post-quench dynamics is set by the energy spectra $\pm R_f(k) $ of $ H_f $. Hence, the momentum-resolved $ g_{tt}(k) $ acts as an indicator of the quantum dynamical propensity, identifying which modes have more active response to the quench.

The features of $g_{tt}(k)$ visible in Fig.~\ref{Fig_gtt} have physical implications. Its vanishing at $k=0,\pi$ is a consequence of a symmetry of the SSH model at these points, where the Bloch Hamiltonian reduces to a form only  proportional to $\sigma_x$ with eigenstates independent of $m$. Consequently, the pre-quench ground state remains an eigenstate of $H_f(k)$ after the quench, yielding a trivial superposition ($|\beta|^2 = 0$ or $1$) with zero energy variance. Consequently, $g_{tt}=0$ at those points. This reflects a symmetry-locking mechanism that suppresses quench-induced energy fluctuations at high-symmetry points.

Meanwhile, the peak location and shape are controlled by the proximity of the initial state to the gap-closing point $m=1$. As seen in Fig.~\ref{Fig_gtt}, when $m_i$ is near 1 (e.g., $m_i=0.9$ or $1.1$), the peak sharpens and shifts toward $|k|=\pi$, reflecting the enhanced geometric sensitivity of near-critical states. The peak height, which follows the sequence $(0.5 \rightarrow 0.1) < (1.1 \rightarrow 2.0) < (1.5 \rightarrow 0.1) < (0.9 \rightarrow 2.0)$, reveals a competition between the quench strength $|m_i-m_f|$ and the initial closeness to criticality. A large parameter jump injects more energy, but an initial state close to the gap closing point ($m_i \approx 1$) amplifies the response by lowering the effective inertia of the system. The maximal response for $m_i=0.9 \rightarrow m_f=2.0$ exemplifies the cooperation of both factors, producing the largest $\Delta E(k)$ across a narrow momentum range.

Therefore, the momentum- and time-resolved map of $ g_{tt}(k) $ provides a geometric preview of the post-quench dynamical landscape. It pinpoints the modes that are most strongly excited ($ |\beta|^2 $ furthest from 0 or 1) and thus carry the largest quantum fluctuations and the highest intrinsic capacity for change (shortest $ \tau_k $). These active modes set the stage for the subsequent evolution, whose detailed temporal structure (e.g., the coherent oscillation period $ \sim 1/\tilde{R}_f(k) $ and the wavepacket spreading time $ \sim \mathrm{Var}(\hat{v}_k)^{-1/2} $) are then governed by the specific post-quench Hamiltonian $ H_f $.

\begin{figure*}[t]
\centering
\includegraphics[width=3.5in, clip]{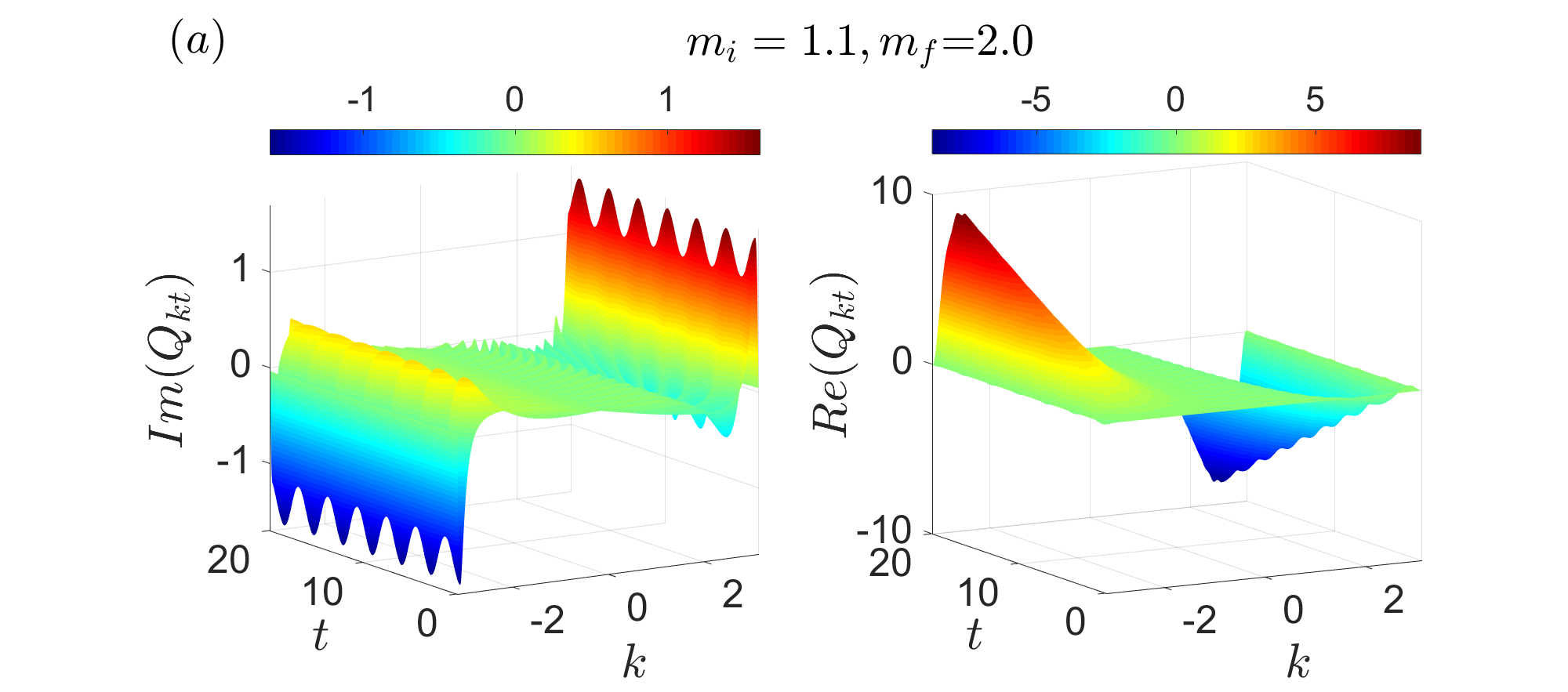}
\includegraphics[width=3.5in, clip]{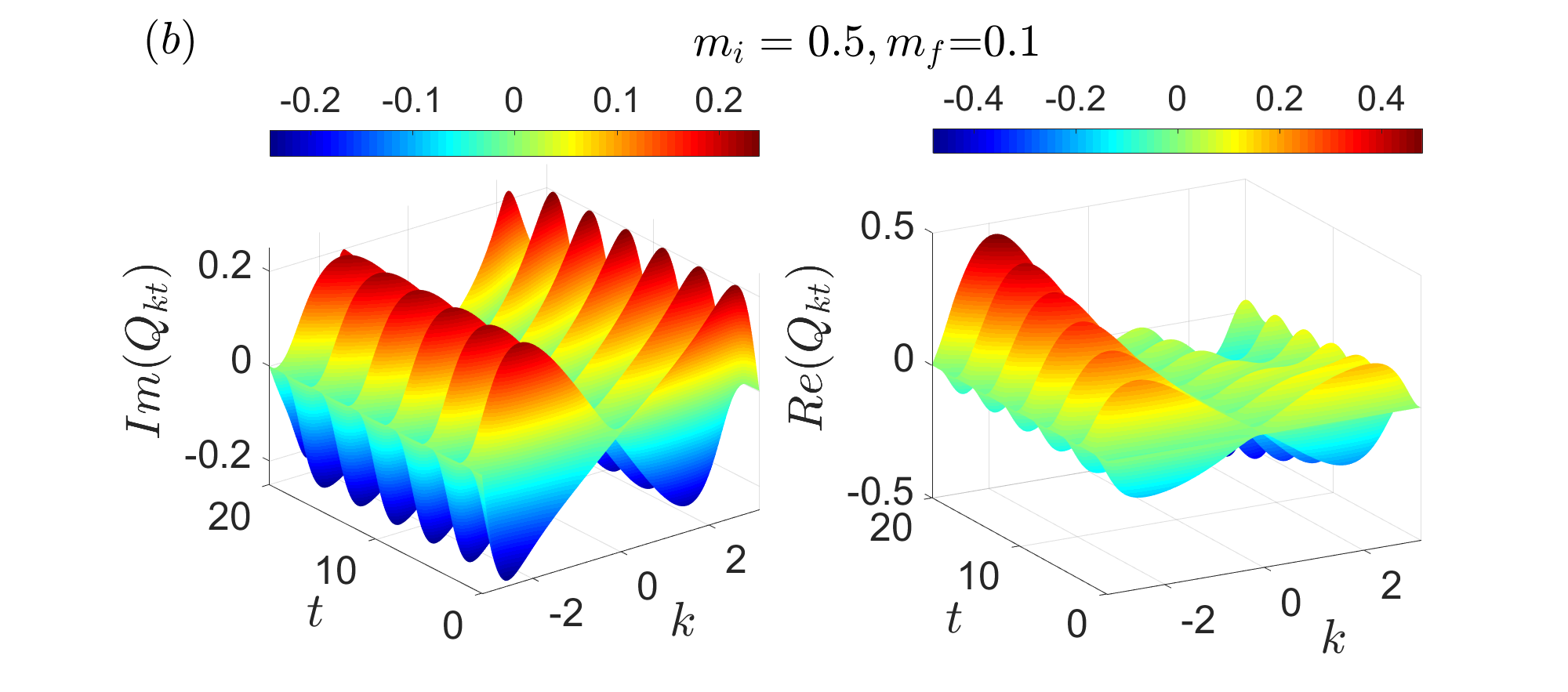}\\
\includegraphics[width=3.5in, clip]{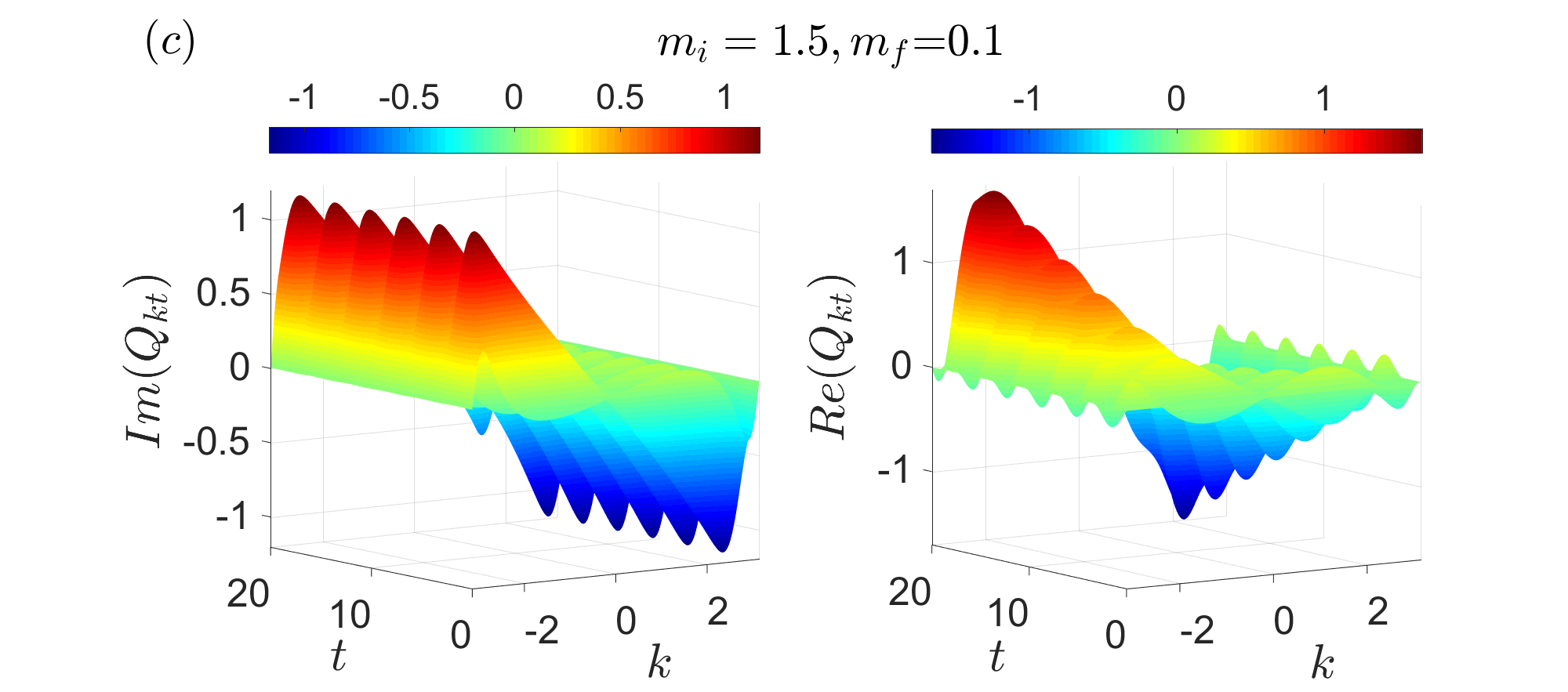}
\includegraphics[width=3.5in, clip]{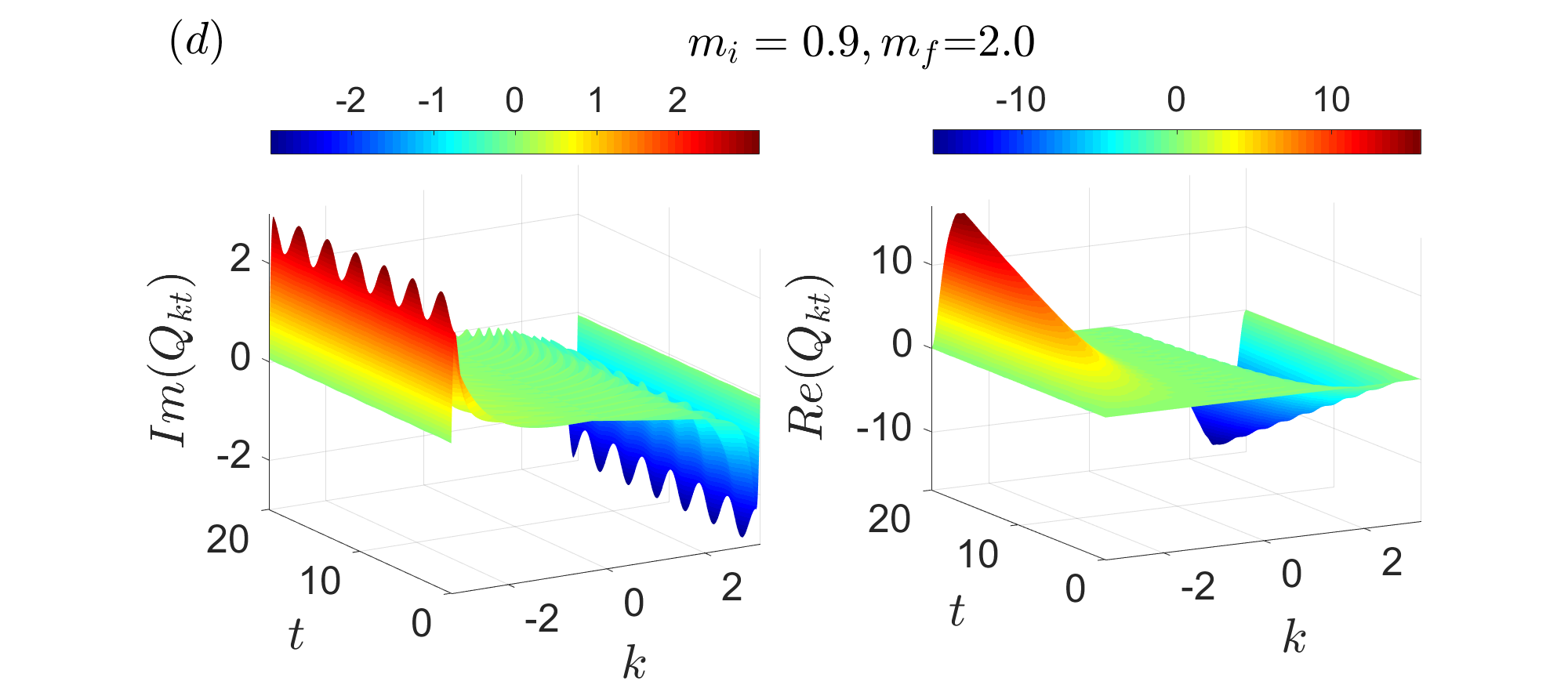}
\caption{Off-diagonal component $Q_{kt}(k,t)$ for the same four quench protocols in Fig.~\ref{Fig_g_kk}.
For each set, the left (right) plot shows the imaginary (real) part.
Here $k\in[-\pi,\pi]$ and $t\in[0,20]$.
The values of $m_i$ and $m_f$ are labeled above each panel.
}
 \label{Fig_Q_kt}
\end{figure*}

\subsection{Off-diagonal component $Q_{kt}$}
The analytical expressions for $Q_{kt}$ in Sec.~\ref{Sec:SSHQkt} shed light on the post-quench geometry on the $(k,t)$ plane. Its dynamics are governed by three constituents: (i) a memory of the initial state geometry ($\propto \mathcal{A}_i$), (ii) coherent oscillations at frequency $2R_f$ driven by the final Hamiltonian, and (iii) for $\operatorname{Re} Q_{kt}$, a secular growth term ($\propto \mathcal{C} t$) that correlates with wavepacket dispersion. This linear-in-time term reflects the dynamical covariance $\operatorname{Cov}(\hat{x},\hat{v}_k)$ between position and group velocity. Since the same coefficient satisfies $\mathcal{C}^2 = \operatorname{Var}(\hat{v}_k)$, the growth of $\operatorname{Re} Q_{kt}$ directly tracks the build-up of position-velocity correlations that precede and accompany the ballistic ($t^2$) spreading seen in $g_{kk}$. Thus, the linear term in $Q_{kt}$ serves as an early-time indicator of the ballistic process.
The prefactor $\mathcal{D}$ shown in Eq.~\eqref{eq:D} is central to the off-diagonal response: Its magnitude scales with $\sqrt{\Delta E^2}$ while its sign, $\operatorname{sign}((m_i-m_f)\sin k)$, imposes the antisymmetry $Q_{kt}(-k,t) = -Q_{kt}(k,t)$ that
ensures the vanishing Brillouin-zone integral $\mathlarger{\int}_{\text{BZ}} (\operatorname{Im} Q_{kt})\dif k = 0$.

Figure~\ref{Fig_Q_kt} displays $\operatorname{Re} Q_{kt}$ and $\operatorname{Im} Q_{kt}$ of the SSH model for the four quench protocols of Fig.~\ref{Fig_g_kk}. All plots obey the required antisymmetry; we describe features in the negative-$k$ half-space.
Figure~\ref{Fig_Q_kt}(a) shows the result of $m_i=1.1 \rightarrow m_f=2.0$. Near $k=-\pi$, $\operatorname{Im} Q_{kt}$ shows narrow negative peaks, because $\mathcal{D}<0$ ($m_i<m_f$) multiplies the large pre-quench Berry connection $\mathcal{A}_i$. Slightly away from the boundary, smaller positive peaks appear as the oscillatory term $-2\mathcal{D}\mathcal{A}_f\sin^2(R_ft)$ becomes significant. $\operatorname{Re} Q_{kt}$ grows almost linearly and smoothly, dominated by $\mathcal{D}\mathcal{C}t$ ($\mathcal{D}\mathcal{C}>0$), signaling rapid dispersion that overwhelms oscillations in this quench above the gap closing point $m=1$.

Figure~\ref{Fig_Q_kt}(b) corresponds to $m_i=0.5 \rightarrow m_f=0.1$. $\operatorname{Im} Q_{kt}$ exhibits broad, rapidly alternating positive/negative stripes along $t$, resulting from small $\mathcal{A}_{i,f}$ and a sign-oscillating factor multiplied by negative $\mathcal{D}$. $\operatorname{Re} Q_{kt}$ shows a linearly decreasing ridge with pronounced oscillatory modulation, indicating strong competition between the dispersion ($\mathcal{D}\mathcal{C}t<0$) and coherence when both initial and final states are well below the gap closing point.

Figure~\ref{Fig_Q_kt}(c) depicts the case $m_i=1.5 \rightarrow m_f=0.1$. $\operatorname{Im} Q_{kt}$ displays only positive peaks (broader in $k$) because $\mathcal{D}>0$ ($m_i>m_f$) and $\mathcal{A}_i-2\mathcal{A}_f\sin^2(R_ft) > 0$. $\operatorname{Re} Q_{kt}$ grows linearly with visible oscillations, resembling panel (b) but with larger amplitude. Crossing the gap closing point does not introduce additional feature as the response remains governed by a competition between quantities from geometry and band structures.

Figure~\ref{Fig_Q_kt}(d) shows $m_i=0.9 \rightarrow m_f=2.0$.
The quench across the gap closing point yields the strongest response. $\operatorname{Im} Q_{kt}$ near $k=-\pi$ shows narrow positive peaks because $\mathcal{A}_i$ is large and positive (since $m_i<1$), while the common prefactor $\mathcal{D}<0$ on the negative-$k$ side.
The initial product $\mathcal{D}\mathcal{A}_i$ thus gives a negative contribution, but the oscillatory term $-2\mathcal{D}\mathcal{A}_f\sin^2(R_f t)$ (with $\mathcal{A}_f<0$) quickly dominates, resulting in a net positive value. The sign reversal relative to panel (a) originates from the opposite sign of $\mathcal{A}_i$ when $m_i$ crosses the gap-closing point, as detailed in Appendix \ref{appc}.
$\operatorname{Re} Q_{kt}$ exhibits the fastest, nearly oscillation-free linear growth among the four cases. This maximum rate follows from the large group-velocity variance $\operatorname{Var}(\hat{v}_k)$, which is maximized when the initial state is close to the gap closing point ($m_i\approx1$) and the quench involves a large $|m_f-m_i|$. Both conditions are satisfied in this case.

The behavior of $Q_{kt}$ provides a refined geometric fingerprint of the quench. Crucially, the sign of $\operatorname{Im} Q_{kt}$ near $k=\pm\pi$ acts as a direct marker of the quench direction ($m_i < m_f$ or $m_i > m_f$), while its amplitude is greatly enhanced by the proximity of the initial state to the gap closing point.
A more detailed analysis is shown in Appendix~\ref{appc}.
In contrast, the secular growth of $\operatorname{Re} Q_{kt}$ primarily reflects the magnitude of wave-packet dispersion. The equilibrium topological invariant given by the winding number itself does not dictate the dynamics; rather, the geometric response is dominated by the interplay of geometric quantities, such as the Berry connection, group velocity, energy variance, and the overlap between the initial and post-quench states. Thus, $Q_{kt}$ serves as a momentum-resolved probe that disentangles the directional, coherent, and ballistic aspects of the post-quench evolution.

\section{Conclusion}\label{Sec:Conclusion}
We have shown that the full quantum geometric tensor provides a complete, momentum-resolved map of post-quench dynamics of the SSH model. The diagonal $Q_{kk}$ gives a metric $g_{kk}$ of the form $g_{kk}^{(0)}+g_{kk}^{(1)}t+g_{kk}^{(2)}t^{2}$. Its long-time ballistic dispersion is driven by a $t^2$ term with the coefficient given by the group-velocity variance, $g_{kk}^{(2)} = \mathrm{Var}(\hat{v}_k)$. The temporal component $Q_{tt}$ coincides with the energy variance $\Delta E^2$. The momentum-space peak of $Q_{tt}$ identifies the modes that evolve most rapidly after the quench, and its sharpness and location are enhanced when the initial state is close to the gap-closing point. The off-diagonal component $Q_{kt}$ is a covariance, exhibiting linear-$t$ term and an oscillatory term in its real part. The imaginary part is a Berry curvature on the $k,t$ plane activated by the quench but does not give rise to additional topological properties for the SSH model.

Collectively, these components establish the QGT as a useful geometric probe for quantum dynamics. The response of the QGT to a quench reveals its dependence on geometric and physical quantities, including the Berry connection, group velocities, and energy variance. While the dynamics of the QGT indicates the changes of the distance between quantum states and local curvature, we have shown that features of the QGT of the SSH model are enhanced by the closeness of the initial state to the gap closing point and the parameter change by the quench. This work complements the idea of using the QGT to decipher quantum dynamics with possible implications for time- or momentum-resolved spectroscopy or control of quantum states out of equilibrium.

\section{Acknowledgements}

H. G. was supported by the Innovation Program for Quantum Science and Technology-National
Science and Technology Major Project (Grant No. 2021ZD0301904) and the National Natural Science
Foundation of China (Grant No. 12447216). X.-Y. H was supported by the National Natural Science
Foundation of China (Grant No. 12405008). C.-C.C. was supported by the NSF (Grant No.
PHY-2310656) for the geometric analysis and the DOE (Grant No. DE-SC0025809) for the physical implications.

Jia-Chen Tang and Xu-Yang Hou contributed equally to this work.

\appendix

\section{$Q_{kk}$ under unitary evolution generated by the same Hamiltonian}\label{app2}
We consider a quantum state that depends on momentum $k$ and time $t$ via
\begin{align}
|\psi(k,t)\rangle = \me^{-\mi E_n(k) t} |n(k)\rangle,
\end{align}
where $|n(k)\rangle$ is a normalized eigenstate of the Hamiltonian $H(k)$ with eigenvalue $E_n(k)$, i.e., $H(k)|n(k)\rangle = E_n(k)|n(k)\rangle$ and $\langle n(k)|n(k)\rangle = 1$. The $kk$-component of the QGT in Eq.~\eqref{eq:QGT} can be obtained by
computing the derivative
\begin{align}
|\partial_k \psi \rangle = \me^{-\mi E_n(k) t} \bigl( -\mi t\, \partial_k E_n(k) | n(k) \rangle + | \partial_k n(k) \rangle \bigr)
\end{align}
and the inner products
\begin{align}
\langle \partial_k \psi | \partial_k \psi \rangle &= t^2 (\partial_k E_n)^2 + \mi t \partial_k E_n \bigl( \langle n | \partial_k n \rangle - \langle \partial_k n | n \rangle \bigr) \notag\\&+ \langle \partial_k n | \partial_k n \rangle, \notag\\
\langle \partial_k \psi | \psi \rangle &= \mi t \partial_k E_n + A^*,
\end{align}
with $A = \langle n | \partial_k n \rangle$. Using $\langle n | \partial_k n \rangle - \langle \partial_k n | n \rangle = 2\mi \operatorname{Im} A$ and $A - A^* = 2\mi \operatorname{Im} A$, we obtain
\begin{align}
Q_{kk} = \langle \partial_k n | \partial_k n \rangle - |A|^2,
\end{align}
which is manifestly independent of time $t$.

Thus, when the evolution is generated by the same Hamiltonian that defines the eigenstate $|n(k)\rangle$, the $kk$-component of the QGT remains constant. In other words, the unitary evolution $\me^{-\mi H t}$ preserves the quantum metric along the $k$-direction. After a quench, however, the evolution is generated by a different Hamiltonian $H_f \neq H_i$, and the time-dependent state is no longer a simple phase-twisted eigenstate. Consequently $g_{kk}$ acquires secular terms that grow linearly and quadratically in time, reflecting the ballistic separation of the states.

\section{Sign of $\mathcal{A}_i$ and its effect on $\operatorname{Im}Q_{kt}$}\label{appc}
The pre-quench Berry connection of the SSH model is given by
\begin{align}
\mathcal{A}_i(k)=\frac{m_i\cos k+1}{2\tilde{R}_i^{2}},\quad
\tilde{R}_i=\sqrt{m_i^{2}+1+2m_i\cos k}\; .
\end{align}
Its sign near the Brillouin-zone boundary $k=\pi$ (where $\cos\pi=-1$) is determined by $1-m_i$:
\begin{enumerate}
  \item If $m_i<1$, then $1-m_i>0$ and $\mathcal{A}_i$ is positive near $k=\pi$.
  \item If $m_i>1$, then $1-m_i<0$ and $\mathcal{A}_i$ is negative near $k=\pi$.
\end{enumerate}
Moreover, when $m_i$ approaches the gap-closing point $m=1$, the denominator $\tilde{R}_i$ vanishes at $k=\pi$, causing $|\mathcal{A}_i|$ to diverge as $\sim 1/|1-m_i|$. Thus, even a small $m_i$ (e.g., $0.9$) can yield a large $\mathcal{A}_i$ if it is close to $1$.

The imaginary part of $Q_{kt}$ is given by Eq.~(\ref{eq:ImQkt}).
For quenches with $m_i<m_f$ (as in both Fig.~\ref{Fig_Q_kt}(a) and (d)) and on the negative-$k$ side ($\sin k<0$), $\mathcal{D}$ is negative. Therefore, the sign of the initial term $\mathcal{D}\mathcal{A}_i$ (at $t=0$) is opposite for the two cases:
\begin{enumerate}
  \item In Fig.~\ref{Fig_Q_kt}(a) ($m_i=1.1>1$) we have $\mathcal{A}_i<0$, so $\mathcal{D}\mathcal{A}_i>0$ (negative times negative).
  \item In Fig.~\ref{Fig_Q_kt}(d) ($m_i=0.9<1$) we have $\mathcal{A}_i>0$, so $\mathcal{D}\mathcal{A}_i<0$ (negative times positive).
\end{enumerate}
Thus, the sign difference explains why the narrow peaks of $\operatorname{Im}Q_{kt}$ near $k=-\pi$ are negative in Fig.~\ref{Fig_Q_kt} (a) but positive in Fig.~\ref{Fig_Q_kt}(d). The subsequent oscillatory term $-2\mathcal{D}\mathcal{A}_f\sin^2(R_ft)$ (with $\mathcal{A}_f<0$ for $m_f>1$) can modify the instantaneous value, but the initial sign set by $\mathcal{A}_i$ remains visible at short times.

%

\end{document}